\documentclass[10pt,twocolumn,a4paper]{article}

\usepackage{graphicx,amsmath,amssymb,amsthm,mathabx,booktabs,txfonts,mathrsfs}
\date{}

\author{Yong Tan\\~\\
\emph{yongtan\_navigation@outlook.com}}
\title{Solve For Shortest Paths Problem Within Logarithm Runtime}

\theoremstyle{plain}

\begin{document}
\maketitle 
\begin{abstract}
The Shortest Paths Problem (SPP) is no longer unresolved. Just for a large scalar of instance on this problem, even we cannot know if an algorithm achieves the computing. Those cutting-edge methods are still in the low performance. If we go to a strategy the best-first-search to deal with computing, it is awkward that the technical barrier from another field\textemdash the database, which with the capable of Online Oriented. In this paper, we will introduce such a synthesis to solve for SPP which comprises various modules therein including such database leads to finish the task in a logarithm runtime.

Through experiments taken on three typical instances on mega-scalar data for transaction in a common laptop, we show off a totally robust, tractable and practical applicability for other projects.
\end{abstract}

\emph{keyword: Shortest Paths Problem; Online Database; Logarithm Runtime.}

\section{Introduction}
The Shortest Paths Problem or SPP is essential one in computer science, meanwhile it is an issue relative to an economic decision so that it possessed of a broad application in the world. This issue was raised in research was in the last century 1950s which has been well known by people. Wherein many researchers involved Shimbel, Bellman, Ford, Moore, Dijkstra, Dantzig and so forth\cite{1} with their seminal works. Today, their devotions are still to affect and support our study as the essential theory and analytic model.

As an analytic model in almost literatures, problem has been often described to reckon the cost of travels with a physical couple of locations as \emph{source} and \emph{destination} in a given map. The object can be considered to comprise two independent systems: (1) is the graph that consists of two kinds of elements the \emph{node} (vertex or endpoint interchangeable) and \emph{edge}; note that among all nodes, an edge is used to join two nodes so each edge allowed with directed property; if we regard a connected relationship as a constraint to traversing on graph, then (2) each edge can get associating to a dimension as \emph{weighted}-edge to quantify this relation, the instance thus becomes a weighted-graph, where the dimension we call \emph{weight} or \emph{cost}; moreover the sum called \emph{total} weight.

When we plan a route getting to a destination from a given source, besides the travels must yielding to those edges which actually maintains a set of constraints to pass; the permutation on such those segments (without repetitions) that composes a path to which those weights incident, we hope the total weight may be conducted into minimum by our scheme. In this case we call that path \emph{optimum}. Apparently, the result is possessed of the combinatorial significance or optimization other than Euclidia's inequality.

Of course, there is a constraint to the data type of weight within many literatures. That is \emph{fixed} as a constant through the whole course. As well, the correspondence of amounts for sources and targets can go to single or many and something else. Yet the fixed weight and single source is only the radical problem imperative to solve for.

\textbf{Related Works.} There are two ways so far well known by us taken to serve calculation. First is by \emph{algebraic} way, which attempts to use connected matrices that load those various relationships relevant to those edges or incident weights in instance. In this field, the prominent figure is Shimbel\cite{1} with complexity $O(n^3)$ for his invented solution. In that way, to get further optimizing by matrix is not easy, for example to read through a $n\times n$ matrix, the complexity can in square $O(n^2)$ technically.

Another is \emph{heuristics} alternative, to which more cult statuses come up they include Dijkstra, Moore, Bellman, Ford, Dantzig and so forth. Their wises almost compass around a functor (in following) proposed by Ford firstly, that is an iterative method is asymptotically to take from the values towards the final convergence\cite{1} upon that \emph{s\textendash t} arc:

\begin{flalign*}
\text{choose an arc }&(s,t)\text{ with } d(t)>d(s)+l(s,t)\\
&\text{ then }d(t)\coloneqq d(s)+l(s,t).
\end{flalign*}

Although the functor has been developed to a few of various forms, but the essential ideas and characteristics cannot be altered as well, only just to adapt different computing wise. Herein we call the functor \emph{Asymptotical Optimizing Functor} (AOF); the arc $(s,t)$ therein, thus called \emph{Shorter Arm} (SA) whereon the node $s$ called \emph{Origin} or origin node, the another $t$ called wild node.

When we study this functor, it is easily to realize to reckon weights in functor is implicitly summable. So if the updates on dependent variant $d(t)$ were discretized to a set of numbers; and if series permutation and every value able to connect with resolving process, then the series would show a \emph{monotonic} characteristics demonstrated on two aspects: (1) existing a sub-path $p(s) \subseteq p(t)$ such that $d(s) \leq d(t)$; (2) the inequality $d^{(i)}(t) \leq d^{(i-1)}(t)$ holds if the superscript $i$ represents different phases in process of update, which would mean that series can \emph{converge} corresponding to the wise of asymptotic update.

Of course, it is explicitly that the \emph{negative} number as weight can damage that characteristics of convergence; i.e. the \emph{decrement} able to circulated in repetition on a circuit line that loads some nodes. Whereon for each $t$, the total weight $d(t)$ might be reduced repeatedly tending to \emph{negative-infinity}. Hence this constraint of \emph{nonnegative} on weight's definition must as a \emph{prerequisite} in our research, therefore, the total weights about a path on the destination $t$ can be allowed to converge at 0.

An algorithm the Bellman-Ford method is the earliest one truly to resolve SPP completely; but, the weakness of which it possessed is noticeable for wasting resources in a want of practical strategy taken to conduct search. Whereas this point, many researchers devoted to resolve this crux. A strategy of \emph{best-first-search} thus arose for this aim. In this way, a set of data structures were invented to manage those dynamic data to gear that strategy, e.g. the \emph{re-distributed} heap raised by Ravindra K. Ahuja, et al (1990)\cite{12}; some conclusions were offered in pledge solving for SPP in linear time by using their wises.

Nonetheless, some dissents you cannot ignore, their attitudes were nothing of kind in optimistic. With their evaluation, those ways even lost to Bellman-Ford's, likewise wearing out resources. According to the report of evaluation that from Cherkassky et al (1993), they threw cold water on readers with trials and reports of theirs in which they tested those up-to-date algorithms\cite{2}.

The worse still if people went on to relax more constraints in their instances. As Zhan and Noon had done those in their traffic project (1996)\cite{2} that of route planning, in which instance was allowed to contain more cycles; the \emph{bi-directional} likewise allowed between two intersections so as to get more approach to duplicate a reality world. 

The scalar for the trial instance contained hundred thousand roads and intersections, their works eventually proved the performance on those algorithms was low, obviously wide off their longing.\\

The \emph{organization} for rest is: next section will in a length to introduce data structure and essential method. Section 3 subsequently interprets an approximate method in \emph{geometric} context. Section 4 is necessary to discuss labeling function. Section 5 introduces strategy, database, algorithms, complexity and interesting themes. Section 6, there are relevant experiments with typical instances, translating those data and optimizing way. Section 7 is the summary.

\section{Preliminary and Notation}
A \emph{weighted} graph always refers to a set $G=\{V,\tau,W\}$, wherein the collection $V$ is a node set $V=1,\ldots, n$ with $n=\vert V\vert$; note that it is permitted for nodes to take a set of natural numbers as alias. 

The Greek letter $\tau$ is the arc collection with $E=\vert\tau\vert$. The connected relationships for a graph on our model depend on those arcs; each can be an ordered pair of nodes like $(s,t)$. Hence, the arc set is able to be divided by divider which the arc's first node equals of. This \emph{quotient} the component $s$ we call \emph{United Subgraph} and, each can be formatted with a Cartesian product $s=R\times L$ in an \emph{one-many} pattern like \emph{astroid} tree. We call the member in set $R$ \emph{root} a singleton; \emph{leaf} in set $L$ that of $m = \vert L\vert$ and $m\geq 0$. If an operation about leaf set, we can call \emph{Relax Leaves}.

The context about weight is not change than that above-mentioned: \emph{nonnegative} and \emph{fixed}; the total weight's function is of \emph{additivity}. We denote a total weight about a path from source to destination $v$ by $\Omega.v$.\\

\textbf{Graph Partition.} There can be a similar partition on node set $V$\textemdash all nodes may be subdivided into a group of components that in a hierarchic society $R=r_1,r_2,\ldots,r_k$ for $R\subseteq V$, where each called \emph{region}. For a node $t\in{r_i}$ such that an arc $\exists(s,t)\in\tau$ is there surely in existence of $s\in{r_{i-1}}$.

That shape of partition demonstrates an isomorphic construct to characterize an asymptotic-grade-form for \emph{source} (in first region) towards to \emph{targets} in \emph{one-many} model. By means of this structure, there is a kind of shortest path\cite{4} which comprises the least number of arcs from source to targets in a morphology model with a geometric-fashion for our computing. Alternatively, the results would likely be constrained within this context.

\section{Geometric SPP}
We will take a length for pseudo code as follows. In advance, there are three arrays taken to tally every node's status through this routine, where each node's natural number alias as index: (1) \emph{Parent Array} $T$, each node's unique immediate predecessor in path as element is there; (2) \emph{Total Weight Array} $W$ loads $\Omega.v$; (3) \emph{Status} $\xi$ deposits the id of region in which node partitioned. All elements are initialized with \emph{zero} before implement routine. In some literatures, the update in that kind of array can be called \emph{labeling} function, especially to arrays $T$ and $W$.\\

\small{@\textbf{Hybrid Algorithm}}
\begin{flushleft}
\small{$R\leftarrow s; ~\xi.s= 1; ~i = 1;$ // $s$ is source and firstly partitioned.\\
~\\
\textbf{Loop} ($\exists{v}$ \emph{for search})\\
00. \textbf{Each} $v$ \textbf{in} $r_i: r_i\in R$ \quad\emph{// as root}\\
01. ~~\textbf{Each} $\ell$ \textbf{in} $L(v)$ \quad\emph{// relax leaves}\\
02. \quad$w = +(\Omega.v,~\omega.(v,\ell));$\\
03. \quad\textbf{If} $\xi.\ell = 0$ \textbf{than}\quad\emph{// $\ell$ is wild}\\
04. \quad\quad$\xi.\ell = i+1$; $R\leftarrow\ell$;\quad\emph{// graph partition.}\\
05. \quad\quad$T.\ell = v$;  $\Omega.\ell = w$;\quad\emph{// first labeling on $\ell$}\\
06. \quad\textbf{Else if} $\xi.\ell>\xi.v$ and $\Omega.\ell > w$ \textbf{than}\\
07. \quad\quad$T.\ell = v$;  $\Omega.\ell = w$;\quad\emph{// AOF.}\\
08. \textbf{End~End}.\\
09. update info. of $r_{i+1}$ in $R$ for next round ($i\coloneqq i+1$).\\
\textbf{Output:} $T, W$.\\
}
\end{flushleft}

In this method, the graph partition impose to relax leaf sets, all leaves in set may be respectively settled in the \emph{previous}, \emph{native} or \emph{next} region contrast to their common root. When graph partition and AOF consorts, the search style can be said in use of the \emph{depth-first-search} tactic like $A^{\ast}$ method. In this way, we no longer concern about the cycle's existence; i.e. the circuit form does not restrain the search totally. 

Alternatively, only \emph{next} leaf, its labeling has chance to obtain optimizing that from its own roots in \emph{previous} region. This measure likes Moore's labeling\cite{1}, over there the region's id yields a shortest-path model in geometric fashion which can take visitor straightly to each target in graph; but it does not resemble Dijkstra's method that needs drawing a non-circuit instance in advance; anyway, both on complexity are as same in $O(E)$ whereto each arc would be screened for precisely once. 

Nonetheless we rather call this method \emph{Hybrid Dijkstra’s Method} or HDM than others, without suspicion, this method can be viewed as derived from Dijkstra's and other's devotions for this problem. 

The results for SPP whose output consists of arrays $T$ and $W$, we can but say \emph{nice} for its accuracy rather than \emph{precise} on account of the geometric context above-mentioned. Of course, if HDM on the non-circuit instance, then it is absolute of optimal result; that proposition is easy to proven.

However, HDM actually takes a work upon initializing our variables and, gives an up bound for those labeling, so that we need a further step to proceed optimization with HDM results as being input entry.

\section{Bellman-Ford Method}
We have to refer to \emph{Bellman-Ford Method}\cite{5} when we discuss resolving SPP.
Particularly, their \emph{Correctness Theorem} has been the standard knowledge appearing in many literatures. To us, the theorem implicates more significances to support our work: Upon to the two labeling arrays $T$ and $W$, if we are dedicated to deal with those SAs, then (1) there is not any circuit incurred in parent-array $T$ through the whole computing course; (2) while no SA for search, the total weights for all targets in array $W$ are optimum, furthermore, which is the signal that can lead program to terminate. 

By that theorem, we can interpolate a module into that code above-mentioned at the 8th line: its task is of detecting the existence of SAs on two directions pointing to either pervious region or native region and, to gather those origins by this way. The new pseudo code is listed in following.\\

\small{@\textbf{Seeking Module}}
\begin{flushleft}
\small{
08. \quad\textbf{if} $\xi.\ell \leq \xi.v$ and $\Omega.\ell > \omega$ and $v\notin C$ \textbf{than}\\
09. \quad\quad$C\leftarrow{v}$};\\
\end{flushleft}

Thus, we have an origin list for executing that asymptotic AOF so that program can go along the veins of $s-t$ in graph to chase many more new origin nodes, meanwhile to update those labels in arrays $T$ or $W$, which resemble to Bellman-Ford wise.

Technically, this new method can spare the computing resources than Bellman-Ford's no longer to look for a few of origins from scratch once again and again; because the labeling operations can be iteratively among origin nodes, none the involving to other wild nodes in graph.

Nonetheless, a crux emerges immediately that an arc is likely to be surveyed or as SA repeatedly over time for many times, i.e. a plenty of repetitive operations would occur for which we cannot assess the relevant scalar of computing. We have to maintain the complexity on this wise up to Bellman-Ford's in $O(nE)$, at most to add a coefficient that of real number and less than 1. Therefore, we call $O(nE)$ Bellman-Ford Benchmark or BFC. 

It is clear for us to go wrong to calculate the dynamic circumstance for labeling function and, the results are not what we hope as well.

\section{Logarithm Method}
In this section, we must solve for two problems. One is what strategy we will take for further AOF; another is to design the database to secure origin’s supplying and at a low cost. 

\emph{The strategy} we will go to is the \emph{best-first-search} raised by Gallo and Pallottino (1988)\cite{2}. Of course, we so far have two choices for the concept of the \emph{best}\textemdash region first or total weight first, but the later is more objective to application. Firstly, we deduce this strategy how to save resources by a gigantic leap over many asymptotic stages, i.e. a quick way upon the convergence for total weight.

For a telling example, consider the case for double origin nodes in a graph as two ends on a string $L=\{{x_1},\ldots,B,\ldots,{y_1}\}$, where nodes $x_1$ and $y_1$ both are origin nodes. We window the string $L$ at endpoint $B$, over there we observe the evolution in string.

We suppose there were a labeling AOF $F_{\in AOF}(x_1)$ work and get off from $x_1$; the progress along the string $L$ binds for the wrong side the origin $y_1$ takes. We can have an analytic model upon the ensuing consequence: (1) While $F(\Omega.B)<\Omega.y_1$, which ruled by the best first clause, this progress might be on and win node $B$ to label it; even the origin $y_1$ could be proven of a \emph{quasi}-origin if $F(\Omega.y_1)<\Omega.y_1$. (2) As well, the tendency from $x_1$ might go checked at window $B$ by $F(\Omega.B)>\Omega.y_1$ and, another AOF would begin at $y_1$ towards to node $B$; hence we can view node $B$ as a border for both kingdoms the $x_1,\ldots$ and the $y_1,\ldots $.

Furthermore, to compare two kinds of nodes underlying \textquotedblleft best\textquotedblright strategy, we have: (1) Although there were likely multiple origins have many wrestles on common border $B$, the count of taking would not yet greater than the \emph{in-degree} on $B$ the number of pipes to $B$ from others. (2) Another kind is that node in kingdom, to which program alternates to change that node’s identity between wild and origin, this change on each arc should cause arc precisely once in use for investigation because of our prerequisite about total weight’s monotonic property. As things stand, the scalar of (1) should be negligible far less than (2) can be ignored. The complexity for such strategy could be estimated by $o(n) + E$, former term therein is the cost in use of database, called \emph{Database Cost} or DC; consequens is \emph{Update Count} or UC.

\subsection{Priority Heap}
For such database, Pape (1974) proposed the conception of \emph{Priority Heap} but his invented model with that heap in an exponential complexity $O(n2^n)$\cite{13}. That construct is actually a dique where all members link one another in a circuit. The visit entrance the minimal item that at the top of heap. When a new item were pushed into this heap, it would be settled at the bottom of heap at first; and then heap should lift up the fresh item, only if it with a key smaller than senior others in heap.

Without suspicion, the interest is easy to fetch the minimal items from top of heap. But to maintain the heap, it reflects a severe problem about cost, while we go in detail to investigate: when lifting an item in heap, the cost could concern the whole construct of heap at the worst case as well as \emph{bubbling} sorting, however for item’s addition or update. It is noticeable that if there a mass of online operations on this heap, the case can in deterioration worse than BFC technically.

\subsection{Bucket Heap}
Another memory technology we will introduce is \emph{Bucket Heap} offered by Dial (1969)\cite{2}. The construct of \emph{bucket} is a hash table is just inverse to labeling array $W$, whose index is the total weight pointing to node. Frankly speaking, this design can effectively avoid the case of which a trifle can cause a global expenditure as well as Pape's heap does; over there, operations of insertion, deletion and update an item in a bucket will within a little cost just as a hash table does.

But there are much more troubles beyond these benefits above-mentioned: (1) How to take a bucket to load a set of \emph{analogous} nodes whose keys are same one and, the number of analogues is only dynamically identified; so the bucket has the necessity to expand to two dimension with each index pointing a nth array. (2) Despite of real number as weight no match hash table, to forecast the size of oneself is a challenge whose size is the product of two variants $n$ and the maximum cost is easy to conduct a wild inflation in memory. 

The complexity for bucket is $O(mb+n(b+C/b))$\cite{2} the coefficient $b$ therein is an unidentified constant; which shows the bucket like a giant monster able to eat up memory swiftly. Excepted the functional unbalance, another challenge is as salient as Bellman-Ford's: for few of origins, program may have to waste resources to look over in a huge bucket; the time resources would in vain to be worn away for this measure.

\subsection{Fibonacci Heap}
A heap with cobweb shape alike, that is \emph{Fibonacci Heap} invented by Johnson, Fredman and Tarjan; by which they attempted to get a balance on every aspect at a low cost. In that structure, an item could be linked with four another ones, moreover, they are constrained one another by their keys in \emph{ascent} at each \emph{column}. On the row, there no sorting on it, nevertheless those items are linked in a cycle too and no visit entrance in that circuit. It can be noted that the heap’s building is rather at will and cheap. An insertion can be in $O(1)$\cite{6}; if further insert a set of items into heap, these fresh items need to be partitioned into a few of columns. Hence they hoped all operations on heap could be retained within logarithm time as in a binary tree.

They alleged program with the support of this heap, it could maintain Dijkstra's method into $O(m+n\log{n})$ (where $m=E$)\cite{1,6}. 

It is obviously for a lattice construct; over there a gigantic cost lurks in such structure which relative to maintain the framework. If we regard a row as a \emph{floor} in a building, these minimal items hence in a dique can be used to construct the \emph{roof} the top of skyscraper (the heap). When program went for a set of minimal items, the case would come worse: if delete an item off roof, how to treat those items that in the incident column and under roof? They have been merging with others in several floors. One wise is to pull them up as bubbles upwards, then all under members would have been treated in two steps\textemdash take apart and re-merge with correlative floors. Apparently this wise means a few of minimal items in roof can lead to a global cost in the heap. Otherwise, it needs another device and additional resources to maintain heap to secure the new entrances for next accessing, whose cost is also a globality for looking through the heap.

It is not curiously in\cite{8} given by Hung and Divoky, they reportedly offered a disappointing conclusion about the performance on this heap for SPP. Technically, Fibonacci Heap as a priority heap, it is scarcely fit to the \textquotedblleft best\textquotedblright strategy that with a cruelly online circumstance.

\subsection{Lizard Entity}
In term of those discussions above, we can outline the requirements on database for manage those origins, which are necessary to deal with four transactions: accessing, building, resorting and gathering. 

Firstly it is identified for us to choose the linked list data structure as those researchers had done above-mentioned. Second goes to a special tree we call \emph{Lizard Entity} or LE, which is a special instance derived from CBST (Compound Binary Search Tree)\cite{9}; in which, we only just addressed a case of all items whose keys different one another.

The CBST comprises two systems that of a Binary Search Tree (BST) and, a Doubly Linked List (DLL) a sorted queue (ARA) in which all items have been queued by their keys in ascent. Three interfaces are thus exposed outside that structure: the minimal item and the maximal item in ARA; the root of BST. This framework and these entrances consort to lead to the accessing cost in: (1) $O(4)$ for \emph{deleting} an item off CBST at worst case, a constant-level; (2) \emph{insertion} may in a low bound $\Theta(\log{n})$; \emph{inquiry} as same as insertion, but to get a minimal item, (3) by the minimal interface in ARA, the cost will be in $O(2)$ rather than Fibonacci heap the low bound within logarithm as same as in a binary tree. With the cursor going along the ARA, program can traverse through this list and able to take the minimal items away within a linear time\cite{9}. 

For building a CBST, the cost will be in $O(n\log{n} + n)$ where the former is sorting cost to yield an ARA in use of \emph{Card Game Sorting Method} or CGSM, another is Pyramid Method or PM\cite{9} to assemble a BST based on ARA. 

Besides accessing and building taken in linear time, a crucial method is re-sorting. The problem would arise when an item's key updates in CBST; which requires an appropriate position in CBST to resettle the item. So there involve two aims: (1) Maintain the logical framework of CBST. (2) Adjust the ARA for a correct accessing. We use a composed method that deleting changed item and then re-inserted, note that the estimations on our accessing operations, the cost may in low bound of $\Theta(4 + \log{n} )$ a real \emph{local} cost.

At the last, we use a DLL as a queue to gather those analogues, for each we call \emph{cousin} and, that list is Cousin List or CL. In fact, the CL is a \emph{third} system independent of CBST but is integrated with CBST by hanging the queue in two models. With the head of CL as an \emph{agency} appearing in two models of CBST, CL can be accessed in convenience or concentratively via CBST model. Eventually, the LE is made up as a priority heap which comprises three systems: BST, ARA and CL.\\

Nonetheless we add a CL to deal with analogues' congestion, those cost of operations are maintained not to change whereto all correspondences on CBST. All above, our solution to SPP is a synthesis integrated of Ford's AOF, Gallo's and Pallottino's strategy and our online database to support; we call the integral method \emph{Contest Algorithm} (CA) which works on the labeling results from HDM as input entries.\\

\small{@\textbf{Contest Algorithm}}
\begin{flushleft}
\small{
/\/* make a LE with array $C$ \/*/\\
make LE: $\mathscr{T}=PM(C);~C.0=0$; // reset counter in $C$\\
~\\
/\/* optimizing course \/*/\\
\textbf{Loop}: $sizeof(\mathscr{T})\neq\varnothing$\\
1. list $C\leftarrow \varepsilon\colon\varepsilon=\emph{getMin}(\mathscr{T})$;\quad\quad// traverse ARA.\\
2. \textbf{Each} node $e\in C$ as root \textbf{do}\quad\quad// relax origin's leaf set.\\
3. \quad\textbf{Each} leaf $\ell\in L(e)$ \textbf{do}\\
4. ~~\quad\textbf{if} $(e,\ell)$ is SA \textbf{than} \emph{update}($T.\ell$, $\Omega.\ell$);\quad// AOF\\
5. \quad\quad\textbf{if} $\ell\in\mathscr{T}$ \textbf{than} \emph{delete}$(\ell, \mathscr{T});$\quad// delete $\ell$ if in LE\\
6. \quad\quad list $\xi\leftarrow\ell$ (for $\ell\notin\xi$);\quad// record $\ell$ if no repetition.\\
7. \textbf{End End}\\
8. \emph{insert}$(x, \mathscr{T})$ (for $x\in\xi$);\\
9. $C.0=0; ~\xi.0=0$ reset counters for next round.\\
}
\end{flushleft}

\textbf{Complexity.} The first phase for building LE takes sorting and linking. At the worst case, the origin nodes in list $C$ about a stochastic permutation, the complexity is none the less obtained in $O(n\log{n} + n)$ because in use of CGSM and PM.

In the second stage, consider the low bound to work resorting able in $\Theta(\log{n} + 4)$, the complexity can be reckoned with $\Theta(n\log{n} + 4n)$. Technologically, the up bound on BST is $O(n^2)$ at worst case. Consider $\vert{C}\vert\ll{n}$ in most search round that of a dynamic circumstance, there appears to be rarely for the risk of worst case occurring, so that we can identify $n\log{n}$ as our DC’s gauge. Thus, we integrate HDM and CA as a whole solution and that further to have an approximate term $O(\lambda{n}\log{n}+2E)$ for runtime complexity, a \emph{quasi-logarithm} time where the coefficient $\lambda$ we call \emph{Harmonic Factor}.

Indeed, it is a matter of fact that the shape of LE out of our conduction in executing AOF phase, so we will not surely connect the LE with a \emph{proper} BST through whole course; i.e. the BST in LE should be taken into account of a \emph{random} tree, moreover, we cannot take a general measurement to estimate the complexity. Yet it may be considered to go to the conception of \emph{average} cost on random tree, that likely in $O(1.39{n}\log{n})$ time, withal the conclusion from empiric spectrum\cite{10}.

\textbf{Instance.} When we consider a \emph{strong} connected graph in $m\approx{n -1}$, such that $E\approx{n^2}$ and $E\gg{n}\log{n}$. Contrast to UC the scale in $O(E)$, apparently, the DC is but a secondary one in search, a small share in whole cost if $n$ is a large number. Conversely for a \emph{weak} connected with $n\gg{m}$ and $\log{n}\gg{m}$, DC should take the main effort of computing.

We beforehand give the parameters which used either for instance or for coming trials.

\begin{center}
Table.1 \small{\textbf{Parameters}}\\
~\\
\renewcommand{\arraystretch}{1.2}
\begin{tabular}{l|lccr}
\hline
\small{\textbf{Inst.}}&$n=$&$n\log n(n^{\ast})$&$E=$&$T_{HDM}$\\
\hline
\small{Comp.}&$2K$&$2.19\cdot 10^4$&$3.998M$&110\\
\hline 
\small{Rand.}&$222K$&$3.94\cdot 10^{6}$&$3.996M$&219\\
\hline
\small{Grid.}&$1M$&$1.99\cdot 10^{8}$&$3.996M$&234\\
\hline
\end{tabular}
\end{center}
Remarks: \emph{\small{(1) with $m=(n-1)$, object \textbf{Comp.} is a complete graph whose each node points to all others; (2) \textbf{Rand.} is random to insert nodes into the leaf set by $m=\log{n}\approx 18$; (3) \textbf{Grid.} a grid instance with $m\leq{4}$ like city layout, on it the number of nodes on row and column, both equal of $1K$. Three $n$s respectively into $2K$, $222K$ and $1M$, thus each quantity of arcs approaches to $4M$, so we can roughly say they go to simple graph with an equivalent scalar. (4) The item $T_{HDM}$} is the practical runtime of executing HDM, for which the \emph{millisecond} (ms) as physical unit for measurement. \footnote{There is a considerable distance among them nevertheless they approach very much. This case should involve to the memory management in OS. We take a 2D array to load an instance and, the second dimension array deposits the leaf set so it could in a consecutive physical space, but the first dimension which loads several subgraphs is just inversely. These delicate differences on visiting could get accumulated to become explicit to show off within big data.}}

Therefore, the equation $\log{n} = m$ can be counted as a threshold applied in our computing, likewise $n\log{n}$ may be as a \emph{benchmark} in use to assess algorithms or instances. In next section, we will analyze the CA by experiments on some typical instances and, take optimization on method. \footnote{Our machine was a laptop with Intel I3 core, 4G memory and Win10 OS. The executive procedure was encoded by C++ at console platform, which from CodeBlocks, version 12.11 (http://www.codeblocks.org/)} \footnote{About our experiment's source code, reader can visit this site\cite{11} to obtain those files or email to author for them.}

\section{Experiments}
About this experiment, the source code has been opened and deposited in that site\cite{11}. Our coming analysis about CA and the efficiency on instances likewise depend on this code and data it took. The weights herein are positive integer with range $[1, 1000]$ and associated to each arc in a random way. Those operations on LE are measured by a uniform way, like traditional wise that counts the number of nodes involved in each practical operation that can be an inquiry, an insertion, a deletion and a building; or other else.

\begin{center}
Table.2 \small{\textbf{ Results.1}}\\
~\\
\renewcommand{\arraystretch}{1.2}
\begin{tabular}{l|rrrr}
\hline
\small{\textbf{Inst.}}&$D$&$Q_{A}$&$Q_{S}$&$Q_{S}/Q_{A}$\\
\hline
\small{Comp.}&7\,577&$3\,996\,001$&10\, 969&0.27\%\\
\hline
\small{Rand.}&504\,719&3\,984\,012&506\,666&12.27\%\\
\hline
\small{Grid.}&1\,323\,416&3\,993\,107&1\,325\,451&33.19\%\\
\hline
\end{tabular}\end{center}
Remarks: \emph{\small{(1) $D$ the sum of total deletions in LE; (2) $Q_{A}$ counts scanning arcs in graph; (3) the total of SAs taken out in whole course is $Q_{S}$.}}

In the table 2, the number of $Q_{A}$ almost equals of parameter $E$ the amount of arcs in graph is practically to tally our forecast in previous section. The ratio $Q_{S}/Q_{v}$ seemingly proves complete instance easily to be optimized. In fact, compare two parameters $Q_{S}$ and $n$, the mean of operations on each node in the complete, the number approach to 6 more than the grid\textquoteright s less than 2.

\begin{center}
Table.3 \small{\textbf{Results.2}}\\
~\\
\renewcommand{\arraystretch}{1.2}
\begin{tabular}{l|rcr}
\hline
\small{\textbf{Inst.}}&$\mathscr{C}=$&$\lambda =\mathscr{C}/n^{\ast}$&$T_{CA}$(ms)\\
\hline
\small{Comp.}&108\,329&$4.94\quad$&203\\
\hline
\small{Rand.}&9\,427\,411&$2.39\quad$&1\,656\\
\hline
\small{Grid.}&36\,778\,016&$1.85\quad$&4\,078\\
\hline
\end{tabular}
\end{center}
Remarks: \emph{\small{(1) $\mathscr{C}$ the total cost of all operations in LE; (2) $T_{CA}$: practical runtime for implement CA}}

In this table, it is interesting of grid instance, whose DC more approach to low bound $n\log{n}$ than others. Contrast to item $T_{HDM}$, the item $T_{CA}$ on three instances demonstrates our analytic about DC and UC, the correlations of their shares in total cost and types of instances. 

Contrast to a similar trial for other algorithms, of complete instance, Mishra et al (2005) reported the optimal result took $2,385$ms on scalar $n=1000$ which was the best score in six executed algorithms. To grid instance, their footsteps balked at the $n=2,500$ and, straightly conceded the unknown risk in bucket wise was the insurmountable hurdle and gave up testing Fibonacci heap too\cite{3}.

In another report for weak connected instance, the scalar on instance contained more than $670K$ nodes and $1.9M$ arcs that from 21 states roads in the United States and used by Zhan and Noon (1996) in their traffic project. In that task to evaluate 15 algorithms in which Bellman-Ford method; Bucket heap and its approximate derivations; and Fibonacci heap and other else were included; yet besides some cautions, Zhan and Noon did not give any data of outcomes about this project finally\cite{2}.

In term of CL data structure, technically, the program can in convenience to exploit that feature of analogues congestion to get the set of minimal items in LE. In general, we can take the whole list by unceasingly reaping the present agency. For instance, given a CL with $k$ ($k>1$) cousins in, one is deleting its agency off LE again-and-again until the whole be reaped, which should in $O(2k)$ time. Another is just to cut the agency of this CL off. Because of that independence, there is merely one time deletion in $O(3)$ time to take the whole list. The efficiency is clearly between both. That evaluation on this optimizing wise is as follows.

\begin{center}
Table.4 \small{\textbf{Optimization Evaluation}}\\
~\\
\renewcommand{\arraystretch}{1.2}
\begin{tabular}{l|rrrr}
\hline
\small{\textbf{Inst.}}&$D^{\prime}(\%)$&$\mathscr{C}^{\prime}(\%)$&$T_{CA}$(ms)&$T^{\prime}$(\%)\\
\hline
\small{Comp.}&31.42&1.38&203&0.00\\
\hline
\small{Rand.}&38.58&6.88&1\,547&6.58\\
\hline
\small{Grid.}&33.79&2.47&3\,656&10.35\\
\hline
\end{tabular}
\end{center}
Remark: \emph{\small{we used \emph{percentage} to measure the degree of improvement on deletion ($D^{\prime}$) , total cost ($\mathscr{C}^{\prime}$) and runtime $T^{\prime}$}}

Of our optimization on LE, it is clearly that not any impaction to UC is merely to reduce the count of deletions in LE. If we take item $T_{CA}$ as an index to evaluate this optimization, there will be: (1) Of the complete, the UC none the less undertakes the main share of overall computing cost, so no change on index. (2) As well on the grid is just inverse to (1), and the two figures the $T^{\prime}$ and the $C^{\prime}$, they are not in proportion. (3) The Random, they are but proportional by its trade-off of threshold.

In a word, this optimization is none the \emph{metamorphosis} just for \emph{constant-factor}, and benefits weak connected instance a lot.

\textbf{Discussion.} A question will be asked that whether it is a necessity for HDM's existence in our solution, whose job is initializing parameters and seems redundant. By our experiments, the data seemingly prove this opinion: in three instances, almost all total weights were updated and, the most improvement beyond $35\%$ in grid instance; the number could be much exaggeration for more than $99\%$ on complete's. On theory, to cancel this step seems to save a cost of $O(E)$ literally.

At a different angle, firstly there is a prudent reason: sometimes the results out from HDM are yet optimal especially to the weight range in some dense and a weak connected graph for computing. Second, out of control on LE at beginning perhaps leads to hardly to assess runtime unless we have found out a wise to keep BST in proper shape at every moment; of course this assertion needs evidences to prove. Third, maybe we need to compare the two stages for some aims.

\section{Summary}
By the support of LE, we totally succeed to solve for SPP in a linear times. The future work will become to study how to deploy the computing to satisfy the various practice requirements. On the other hand, we need to study LE and perfect it for working in a more steady state.

\end{document}